# Modeling Concentration Profiles in Electrolytes by Solving 3-D Poisson-Nernst-Planck Equations via Finite Difference Method

*Yitao He[a,*], Dan Zhao[a]*

[a] *Department of New Energy Science and Engineering, School of Energy and Environment, Anhui University of Technology, Ma'anshan, China*

Corresponding e-mail:
* Yitao He: heyitao@ahut.edu.cn; h-yitao@qq.com

**Abstract**

The Poisson-Nernst-Planck (PNP) equations are fundamental for modeling ion transport in electrochemical systems, capturing the intricate interplay of concentration gradients, electric fields, and ion fluxes essential for applications such as energy storage devices and other electrochemical devices. This study introduces a refined numerical framework employing the finite difference method to solve the 3-D PNP equations, enabling precise simulation of ion concentration distributions under realistic boundary conditions and applied electric fields. By rigorously addressing stability criteria and integrating advanced boundary constraints, including the Butler-Volmer equation for surface reactions, the model provides comprehensive insights into ion dynamics, particularly near electrode surfaces where electric field and reaction effects dominate. This framework significantly enhances traditional PNP modeling by accommodating varied boundary conditions, diffusion anisotropy, and complex electrochemical environments, offering a robust tool for investigating electrochemical processes and guiding the design of advanced electrochemical systems.

**1 Introduction**

The concentration distribution is a critical factor in electrochemical systems, such as batteries, as it directly influences ion behavior, current flow, electric fields, and electrochemical reactions. The Poisson-Nernst-Planck (PNP) equations are commonly used to model current density or concentration distributions.[1-3] However, the specific conditions under which these equations are solved are often not clearly defined in the existing literature. In this study, we frame the PNP equations within a well-defined mathematical context and provide a step-by-step guide for solving them. Additionally, we discuss the stability conditions necessary for obtaining accurate three-dimensional (3-D) concentration distributions.

In fact, in 1962, Newman and Tobias[4] simulated the current density in porous electrodes by solving the 1-D Nernst-Planck (N-P) equation under potentiostatic or non-polarization conditions.

But the demand for 2-D or 3-D distribution was continuing presence. Subsequently, many researchers have tried to solve the higher-order PNP equation. Among them, someone used finite difference method (FDM) because of positivity-preserving properties[5], high-resolution discretization[6], and other advantages. For examples, Liu and Wang[7] developed an FDM approach that satisfies the free energy constraints to solve the 2-D PNP equation. Xu et al.[8] proposed a Debye-Hückel-modified PNP equation solved by FDM, and compared it to WKB approximation. However, the stability conditions of FDM when solving the PNP equation still require further investigation, particularly in the context of electrochemical systems. This is because a successful solution is not always achievable in complex systems with numerous electrochemical factors.

In this work, we built a 3-D model according to PNP equation and FDM first, and linked it with Butler-Volmer (B-V) equation and other electrochemical conditions to achieve concentration distribution in 3-D space and on electrode surface. Importantly, the stability conditions and other compatible conditions for electrochemical systems have been proposed.

## 2 Procedures of Solving Equations and Meaning in Electrochemical Systems

Because the finite difference method will be discussed, we need to briefly introduce the whole process of the method like Figure 1.

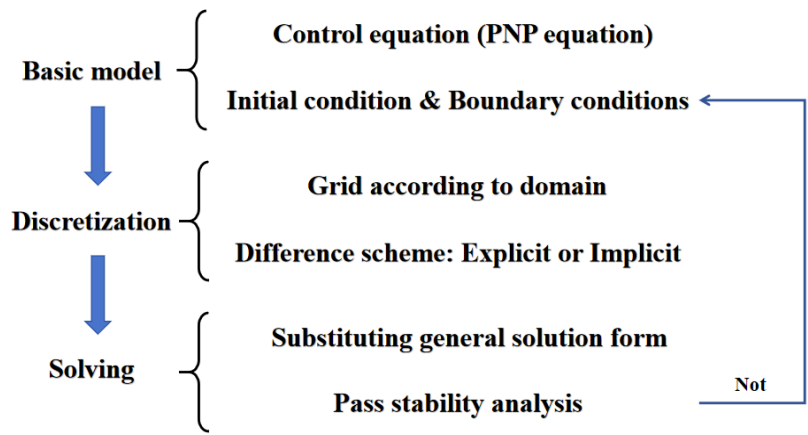

Figure 1. A general process of solving PDEs using the finite difference method.

### 2.1 One Ion Species Model

#### 2.1.1 Physical Descriptions of the Model

The 3-D domain depicted in Figure 2 outlines the spatial configuration required to model ionic transport and reactions within an electrochemical system. The domain is bounded by finite-layer conditions, where no flux is allowed across the lateral boundaries ($x_{left}$, $x_{right}$, $y_{left}$, $y_{right}$, $z_{top}$), effectively isolating the system from external influences. At the electrode surface ($z_{bottom}$), the B-V equation is applied to capture the electrochemical reaction kinetics. This boundary condition governs the ion exchange between the electrode and the electrolyte, driven by the electric field and reaction rate, thereby simulating ion accumulation or depletion near the electrode surface. Together, these conditions provide a realistic framework for studying the coupled effects of ion transport, electric potential distribution, and surface reactions.

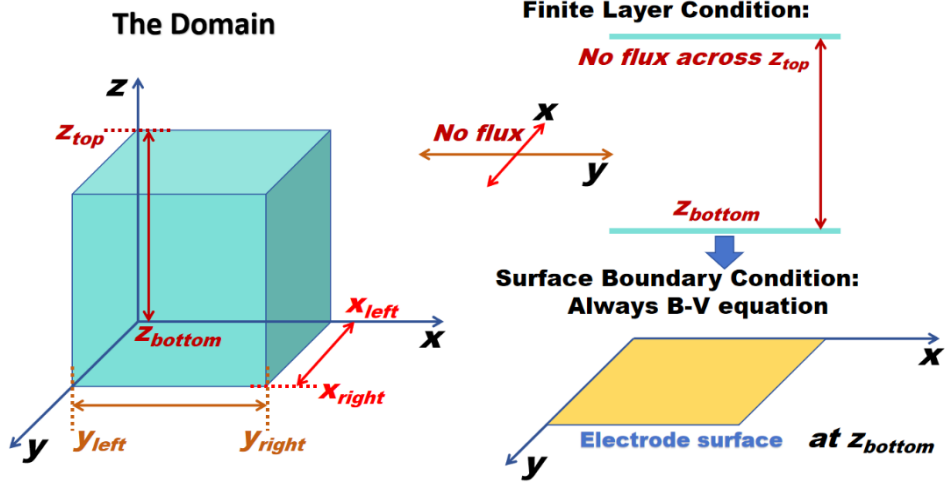

Figure 2. Scheme of the 3-D domain in model.

To establish a robust modeling foundation, we first consider a single-ion system. This approach simplifies the system, allowing a focused analysis of transport and reaction dynamics. By isolating a single ionic species, the influence of the electric field and reaction kinetics on concentration profiles can be clearly understood and validated. Additionally, starting with a single-ion system facilitates the accurate implementation of boundary conditions, such as the B-V equation, which often applies specifically to cations at the electrode surface. This incremental modeling strategy provides essential insights and lays the groundwork for extending the model to multi-ion systems while ensuring numerical stability and computational efficiency.

The assumptions made in this mode include:
(1) Electrolyte homogeneity: The electrolyte is assumed to be homogeneous, with uniform initial concentrations of cations and anions;
(2) Electrostatic potential: The electric field within the electrolyte is derived from the Poisson equation, considering the charge density distribution of the ions;
(3) Ion transport mechanisms: The transport of ions is modeled as a combination of diffusion due to concentration gradients, migration or drift caused by electric field, and electrochemical reaction at electrode surface for cation;
(4) No ion flux at finite-layer boundaries: The bulk electrolyte boundary and lateral boundaries enforce a no-flux condition, ensuring ion conservation;
(5) Neglect of nonlinearities: Effects such as ion crowding or quantum mechanical interactions are not included in the current model to maintain computational feasibility.

**2.1.2 Building Basic Model**

3-D PNP model consists of N-P equation and Poisson equation. For a single ion species $\alpha$, the 3-D N-P equation is[9]:

$$\mathbf{J}_\alpha = -D_\alpha \nabla c_\alpha - \frac{D_\alpha z_\alpha F}{RT} c_\alpha \nabla \Phi \quad (1)$$

where $\mathbf{J}_\alpha = \left( J_{\alpha_x}, J_{\alpha_y}, J_{\alpha_z} \right)$ is the flux vector in 3D; $D_\alpha$ is the diffusion coefficient for ion species $\alpha$, and it will be different along with different directions; $c_\alpha(x,y,z,t)$ is the concentration of ion species $\alpha$; $z_\alpha$ is the charge number of ion species $\alpha$; $F$ is the Faraday constant; $R$ is the universal gas constant; $T$ is the absolute temperature; $\Phi(x,y,z)$ is the electrostatic potential;

$\nabla \Phi = \left( \dfrac{\partial \Phi}{\partial x}, \dfrac{\partial \Phi}{\partial y}, \dfrac{\partial \Phi}{\partial z} \right)$ is the gradient of the potential, also known as the electric field. The electric field change in all three directions will be considered.

The domains are:

$$\begin{cases} x_{left} \leq x \leq x_{right} \\ y_{left} \leq y \leq y_{right} \\ z_{bottom} \leq z \leq z_{top} \\ 0 \leq t \leq t_F \end{cases} \quad (2)$$

In 3-D, the components of the flux are:

$$\begin{cases} J_{\alpha_x} = -D_{\alpha_x} \dfrac{\partial c_\alpha}{\partial x} - \dfrac{D_{\alpha_x} z_\alpha F}{RT} c_\alpha \dfrac{\partial \Phi}{\partial x} \\ J_{\alpha_y} = -D_{\alpha_y} \dfrac{\partial c_\alpha}{\partial y} - \dfrac{D_{\alpha_y} z_\alpha F}{RT} c_\alpha \dfrac{\partial \Phi}{\partial y} \\ J_{\alpha_z} = -D_{\alpha_z} \dfrac{\partial c_\alpha}{\partial z} - \dfrac{D_{\alpha_z} z_\alpha F}{RT} c_\alpha \dfrac{\partial \Phi}{\partial z} \end{cases} \quad (3)$$

which describes the ion flux in the *x*, *y*, and *z* directions due to diffusion and electric field-driven migration.

The Poisson equation in 3-D is[10]:

$$-\nabla \cdot (\epsilon \nabla \Phi) = F z_\alpha c_\alpha \quad (4)$$

where $\epsilon$ is the dielectric constant;

In 3-D, the Laplacian $\nabla^2 \Phi$ is:

$$\nabla^2 \Phi = \dfrac{\partial^2 \Phi}{\partial x^2} + \dfrac{\partial^2 \Phi}{\partial y^2} + \dfrac{\partial^2 \Phi}{\partial z^2} \quad (5)$$

Thus, the Poisson equation becomes:

$$-\left( \dfrac{\partial^2 \Phi}{\partial x^2} + \dfrac{\partial^2 \Phi}{\partial y^2} + \dfrac{\partial^2 \Phi}{\partial z^2} \right) = \dfrac{F z_\alpha c_\alpha}{\epsilon} \quad (6)$$

The N-P and Poisson are parabolic and elliptic equations, respectively. In order to solve this system of two equations, we need one initial condition (IC) and six boundary conditions (BCs) for former, and other six BCs for latter. Actually, no IC for Poisson equation indicates that the electric field is independent of time.

For Nernst-Planck equation, we need an initial condition for the concentration $c_\alpha$ at $t=0$:

$$c_\alpha(x,y,z,t=0) = f(x,y,z) \quad (7)$$

We assume the ions in electrolyte (1 M concentration) will uniformly disperse on the electrode surface at the beginning, so

$$c_\alpha(x,y,z,t=0) = 1 \quad (8)$$

**2.1.3 Discretizing Model Equations**

2.1.3.1 Discretization of the 3-D Domain

For a cubic domain with dimensions $L_x \times L_y \times L_z$, we define a grid of points:

Let $Nx, Ny, Nz$ be the number of grid points in the x, y, and z directions, respectively.

The spatial step sizes will be $\Delta x = \dfrac{L_x}{N_x - 1}, \Delta y = \dfrac{L_y}{N_y - 1}, \Delta z = \dfrac{L_z}{N_z - 1}$. The time step will be $\Delta t = \dfrac{t_F}{N}$ for the time domain $(0, t_F)$.

2.1.3.2 Discretization of the Nernst-Planck Equation:

We denote the approximate solution:

$$J^n_{\alpha_{r,j,s}} \approx J(x_r, y_j, z_s, t_n) \quad (9)$$

where $\begin{cases} 0 \leq r \leq N_x - 1 \\ 0 \leq j \leq N_y - 1 \\ 0 \leq s \leq N_z - 1 \\ 0 \leq n \leq N \end{cases}$.

The explicit difference scheme of Nernst-Planck equation in 3-D can be written as:

$$\begin{cases} J_{\alpha_x}(r,j,s) = -D_{\alpha_x} \dfrac{c_\alpha(r+1,j,s) - c_\alpha(r-1,j,s)}{2\Delta x} - \dfrac{D_{\alpha_x} z_\alpha F}{RT} c_\alpha(r,j,s) \dfrac{\Phi(r+1,j,s) - \Phi(r-1,j,s)}{2\Delta x} \\ J_{\alpha_y}(r,j,s) = -D_{\alpha_y} \dfrac{c_\alpha(r,j+1,s) - c_\alpha(r,j-1,s)}{2\Delta y} - \dfrac{D_{\alpha_y} z_\alpha F}{RT} c_\alpha(r,j,s) \dfrac{\Phi(r,j+1,s) - \Phi(r,j-1,s)}{2\Delta y} \\ J_{\alpha_z}(r,j,s) = -D_{\alpha_z} \dfrac{c_\alpha(r,j,s+1) - c_\alpha(r,j,s-1)}{2\Delta z} - \dfrac{D_{\alpha_z} z_\alpha F}{RT} c_\alpha(r,j,s) \dfrac{\Phi(r,j,s+1) - \Phi(r,j,s-1)}{2\Delta z} \end{cases}$$

(19)

Then, the concentration update for $c_\alpha$ is governed by the continuity equation:

$$\frac{\partial c_\alpha}{\partial t} = -\nabla \cdot \mathbf{J}_\alpha \quad (10)$$

The equation can be discretized as:

$$\frac{c_\alpha(r,j,s,n+1) - c_\alpha(r,j,s,n)}{\Delta t} = -\frac{J_{\alpha_x}(r+1,j,s) - J_{\alpha_x}(r-1,j,s)}{2\Delta x} - \frac{J_{\alpha_y}(r,j+1,s) - J_{\alpha_y}(r,j-1,s)}{2\Delta y} - \frac{J_{\alpha_z}(r,j,s+1) - J_{\alpha_z}(r,j,s-1)}{2\Delta z}$$

(11)

This equation quantifies how the flux or concentration changes with the time in the 3-D domain. Then, the discretization of diffusion term can be written as:

$$c_{r,j,s}^{n+1} = c_{r,j,s}^n + \Delta t \left( D_{\alpha_x} \frac{c_{r+1,j,s}^n - 2c_{r,j,s}^n + c_{r-1,j,s}^n}{\Delta x^2} + D_{\alpha_y} \frac{c_{r,j+1,s}^n - 2c_{r,j,s}^n + c_{r,j-1,s}^n}{\Delta y^2} + D_{\alpha_z} \frac{c_{r,j,s+1}^n - 2c_{r,j,s}^n + c_{r,j,s-1}^n}{\Delta z^2} \right)$$

(12)

2.1.3.3 Discretization of the Poisson Equation:

The Poisson equation in 3-D has been obtained, which can be discretized using central differences for the second derivatives:

$$-\frac{\Phi(i+1,j,s) - 2\Phi(i,j,s) + \Phi(i-1,j,s)}{\Delta x^2} - \frac{\Phi(i,j+1,s) - 2\Phi(i,j,s) + \Phi(i,j-1,s)}{\Delta y^2} - \frac{\Phi(i,j,s+1) - 2\Phi(i,j,s) + \Phi(i,j,s-1)}{\Delta z^2} = \frac{Fz_\alpha c_\alpha(i,j,s)}{\epsilon}$$

(13)

Rearranging for $\Phi(i,j,s)$,

$$\Phi(i,j,s) = \frac{\Delta x^2 \Delta y^2 \Delta z^2}{\Delta x^2 \Delta y^2 + \Delta y^2 \Delta z^2 + \Delta z^2 \Delta x^2} \left( \frac{Fz_\alpha c_\alpha(i,j,s)}{\epsilon} + \frac{\Phi(i+1,j,s) + \Phi(i-1,j,s)}{\Delta x^2} + \frac{\Phi(i,j+1,s) + \Phi(i,j-1,s)}{\Delta y^2} + \frac{\Phi(i,j,s+1) + \Phi(i,j,s-1)}{\Delta z^2} \right)$$

(14)

This equation quantifies how the potential distribution in the 3-D domain, or we can say that this equation describes how the spatial distribution of charges creates an electric potential field in 3-D space.

**2.1.4 Stability Analysis**

2.1.4.1 For Nernst-Planck Equation
*Diffusion term*
We apply Fourier stability analysis to the explicit difference scheme of Nernst-Planck equation. Assuming the general solution form can be written as a sum of Fourier modes:

$$c_\alpha^n(r,j,s) = \hat{c}_\alpha^n e^{i(k_x x_r + k_y y_j + k_z z_s)} \quad (15)$$

where $\hat{c}_\alpha^n$ is the Fourier coefficient of the mode; $k_x, k_y, k_z$ is the wavenumber in the x, y, z-direction; $x_r = r\Delta x, y_j = j\Delta y, z_s = s\Delta z$ is the grid point in the x, y, z-direction. Substituting this equation into the discretized Eq.(12). The diffusion term for can be written as:

$$\begin{cases} c_\alpha(r+1,j,s) - 2c_\alpha(r,j,s) + c_\alpha(r-1,j,s) = \hat{c}_\alpha^n \left(e^{ik_x\Delta x} + e^{-ik_x\Delta x} - 2\right) \\ c_\alpha(r,j+1,s) - 2c_\alpha(r,j,s) + c_\alpha(r,j-1,s) = \hat{c}_\alpha^n \left(e^{ik_y\Delta y} + e^{-ik_y\Delta y} - 2\right) \\ c_\alpha(r,j,s+1) - 2c_\alpha(r,j,s) + c_\alpha(r,j,s-1) = \hat{c}_\alpha^n \left(e^{ik_z\Delta z} + e^{-ik_z\Delta z} - 2\right) \end{cases} \quad (16)$$

Using the identity $e^{ik_{x,y,z}\Delta x,\Delta y,\Delta z} + e^{-ik_{x,y,z}\Delta x,\Delta y,\Delta z} = 2\cos(k_{x,y,z}\Delta x,\Delta y,\Delta z)$, these simplify to:

$$\begin{cases} c_\alpha(r+1,j,s) - 2c_\alpha(r,j,s) + c_\alpha(r-1,j,s) = \hat{c}_\alpha^n \cdot 2\left(\cos(k_x\Delta x) - 1\right) \\ c_\alpha(r,j+1,s) - 2c_\alpha(r,j,s) + c_\alpha(r,j-1,s) = \hat{c}_\alpha^n \cdot 2\left(\cos(k_y\Delta y) - 1\right) \\ c_\alpha(r,j,s+1) - 2c_\alpha(r,j,s) + c_\alpha(r,j,s-1) = \hat{c}_\alpha^n \cdot 2\left(\cos(k_z\Delta z) - 1\right) \end{cases} \quad (17)$$

Therefore, for the diffusion part, the update equation becomes:

$$\hat{c}_\alpha^{n+1} = \hat{c}_\alpha^n \left[1 + 2\Delta t \left(D_{\alpha_x}\frac{\cos(k_x\Delta x)-1}{\Delta x^2} + D_{\alpha_y}\frac{\cos(k_y\Delta y)-1}{\Delta y^2} + D_{\alpha_z}\frac{\cos(k_z\Delta z)-1}{\Delta z^2}\right)\right] \quad (18)$$

Then, the amplification factor considering diffusion term is:

$$\lambda_{diffusion} = 1 - 2\mu_x(1-\cos(k_x\Delta x)) - 2\mu_y(1-\cos(k_y\Delta y)) - 2\mu_z(1-\cos(k_z\Delta z)) \quad (19)$$

where

$$\mu_x = \frac{D_{\alpha_x}\Delta t}{\Delta x^2}, \quad \mu_y = \frac{D_{\alpha_y}\Delta t}{\Delta y^2}, \quad \mu_z = \frac{D_{\alpha_z}\Delta t}{\Delta z^2}$$

*Migration term*

For the migration term containing the electric field, the Fourier mode analysis introduces terms that contribute to the electric potential Φ. And we assume that Φ is independent of time.

In the explicit finite difference discretization, the migration term is approximated as:

$$\begin{cases} J_{migration,x} = -\left(\frac{D_{\alpha_x}z_\alpha F}{RT}\right) \cdot \frac{c_\alpha(r+1,j,s)\Phi(r+1,j,s) - c_\alpha(r-1,j,s)\Phi(r-1,j,s)}{2\Delta x} \\ J_{migration,y} = -\left(\frac{D_{\alpha_y}z_\alpha F}{RT}\right) \cdot \frac{c_\alpha(r,j+1,s)\Phi(r,j+1,s) - c_\alpha(r,j-1,s)\Phi(r,j-1,s)}{2\Delta y} \\ J_{migration,z} = -\left(\frac{D_{\alpha_z}z_\alpha F}{RT}\right) \cdot \frac{c_\alpha(r,j,s+1)\Phi(r,j,s+1) - c_\alpha(r,j,s-1)\Phi(r,j,s-1)}{2\Delta z} \end{cases} \quad (20)$$

For stability analysis, assume the solution can be written as a Fourier mode:

$$c_\alpha(r,j,s,t) = \hat{c}_\alpha(t)e^{i(k_x x_r + k_y y_j + k_z z_s)} \quad (21)$$

where $\hat{c}_\alpha(t)$ is the Fourier coefficient. Substituting the Fourier mode into the migration term. The Fourier transform of the concentration is:

$$c_\alpha(r+1,j,s) = \hat{c}_\alpha(t)e^{ik_x(x_r+\Delta x)} = \hat{c}_\alpha(t)e^{ik_x x_r}e^{ik_x\Delta x} \quad (22)$$

Similarly,

$$c_\alpha(r-1, j, s) = \hat{c}_\alpha(t)e^{ik_x x_r}e^{-ik_x \Delta x} \quad (23)$$

The migration term in the Fourier domain becomes:

$$\begin{cases} J_{migration,x} = -\left(\dfrac{D_{\alpha_x} z_\alpha F}{RT}\right)\hat{c}_\alpha(t)e^{ik_x x_r} \cdot \dfrac{e^{ik_x \Delta x}\Phi(r+1,j,s) - e^{-ik_x \Delta x}\Phi(r-1,j,s)}{2\Delta x} \\[2mm] J_{migration,y} = -\left(\dfrac{D_{\alpha_y} z_\alpha F}{RT}\right)\hat{c}_\alpha(t)e^{ik_y y_j} \cdot \dfrac{e^{ik_y \Delta y}\Phi(r,j+1,s) - e^{-ik_y \Delta y}\Phi(r,j-1,s)}{2\Delta y} \\[2mm] J_{migration,z} = -\left(\dfrac{D_{\alpha_z} z_\alpha F}{RT}\right)\hat{c}_\alpha(t)e^{ik_z z_s} \cdot \dfrac{e^{ik_z \Delta z}\Phi(r,j,s+1) - e^{-ik_z \Delta z}\Phi(r,j,s-1)}{2\Delta z} \end{cases} \quad (24)$$

Using Euler's identity ($e^{i\theta} + e^{-i\theta} = 2\cos(\theta)$), and the migration term simplifies to:

$$\begin{cases} J_{migration,x} = -\left(\dfrac{D_{\alpha_x} z_\alpha F}{RT}\right)\hat{c}_\alpha(t)e^{ik_x x_r} \cdot \dfrac{\sin(k_x \Delta x)}{\Delta x}\Phi \\[2mm] J_{migration,y} = -\left(\dfrac{D_{\alpha_y} z_\alpha F}{RT}\right)\hat{c}_\alpha(t)e^{ik_y y_j} \cdot \dfrac{\sin(k_y \Delta y)}{\Delta y}\Phi \\[2mm] J_{migration,z} = -\left(\dfrac{D_{\alpha_z} z_\alpha F}{RT}\right)\hat{c}_\alpha(t)e^{ik_z z_s} \cdot \dfrac{\sin(k_z \Delta z)}{\Delta z}\Phi \end{cases} \quad (25)$$

The amplification factor for the migration term combines the contributions from the all the directions, which can be expressed as:

$$\lambda_{migration} = 1 - \mu_{E_x}\sin(k_x \Delta x) - \mu_{E_y}\sin(k_y \Delta y) - \mu_{E_z}\sin(k_z \Delta z) \quad (26)$$

where $\mu_{E_x} = \dfrac{D_{\alpha_x} z_\alpha F \Delta t}{RT \Delta x}, \mu_{E_y} = \dfrac{D_{\alpha_y} z_\alpha F \Delta t}{RT \Delta y}, \mu_{E_z} = \dfrac{D_{\alpha_z} z_\alpha F \Delta t}{RT \Delta z}$.

*Stability conditions*

Then, the total amplification factor $\lambda$ for the scheme is:

$$\lambda = 2 - 2\mu_x(1-\cos(k_x \Delta x)) - 2\mu_y(1-\cos(k_y \Delta y)) - 2\mu_z(1-\cos(k_z \Delta z)) - \mu_{E_x}\sin(k_x \Delta x) - \mu_{E_y}\sin(k_y \Delta y) - \mu_{E_z}\sin(k_z \Delta z)$$

(27)

According to $|\lambda| \leq 1$ the condition can be expressed as:

a) Case 1: $1 \geq \lambda \geq 0$
The stability condition for this case is:

$$\mu_x + \mu_y + \mu_z \leq \frac{1}{4} + \frac{1}{4}(\mu_{E_x} + \mu_{E_y} + \mu_{E_z}) \quad (28)$$

b) Case 2: $-1 \leq \lambda \leq 0$
The stability condition for this case is:

$$\mu_x + \mu_y + \mu_z \leq \frac{3}{4} + \frac{1}{4}(\mu_{E_x} + \mu_{E_y} + \mu_{E_z}) \quad (29)$$

Then,

$$\mu_x + \mu_y + \mu_z \leq \min\left(\frac{1}{4} + \frac{1}{4}(\mu_{E_x} + \mu_{E_y} + \mu_{E_z}), \frac{3}{4} + \frac{1}{4}(\mu_{E_x} + \mu_{E_y} + \mu_{E_z})\right) \quad (30)$$

2.1.4.2 For Poisson Equation

The discretized form of the Poisson equation in the x-, y-, and z-directions is:

$$\frac{\Phi(r+1,j,s) - 2\Phi(r,j,s) + \Phi(r-1,j,s)}{\Delta x^2} + \frac{\Phi(r,j+1,s) - 2\Phi(r,j,s) + \Phi(r,j-1,s)}{\Delta y^2} + \frac{\Phi(r,j,s+1) - 2\Phi(r,j,s) + \Phi(r,j,s-1)}{\Delta z^2} = -\frac{\rho(r,j,s)}{\epsilon}$$

(31)

where $\rho(r,j,s) = z_\alpha F c_\alpha(r,j,s)$.

Assume the solution for $\Phi(r,j,s)$ can be written as a Fourier mode:

$$\Phi(r,j,s) = \hat{\Phi} e^{i(k_x' x_r + k_y' y_j + k_z' z_s)} \quad (32)$$

where

$\hat{\Phi}$ is the Fourier coefficient; $k_x', k_y', k_z'$ are the wavenumbers in the *x*-, *y*-, and *z*-directions, respectively. Substituting the Fourier mode into the discretized Poisson equation for each spatial direction gives:

$$\frac{e^{ik_x'(x_r+\Delta x)} - 2e^{ik_x' x_r} + e^{ik_x'(x_r-\Delta x)}}{\Delta x^2} + \frac{e^{ik_y'(y_j+\Delta y)} - 2e^{ik_y' y_j} + e^{ik_y'(y_j-\Delta y)}}{\Delta y^2} + \frac{e^{ik_z'(z_s+\Delta z)} - 2e^{ik_z' z_s} + e^{ik_z'(z_s-\Delta z)}}{\Delta z^2} = -\frac{\rho(r,j,s)}{\epsilon}$$

(33)

After using the Euler's identity,

$$\frac{2\cos(k_x' \Delta x) - 2}{\Delta x^2} + \frac{2\cos(k_y' \Delta y) - 2}{\Delta y^2} + \frac{2\cos(k_z' \Delta z) - 2}{\Delta z^2} = -\frac{\rho(r,j,s)}{\epsilon} \quad (34)$$

The amplification factor for the Poisson equation can be written as:

$$\lambda_\Phi = -2\left(\frac{1-\cos(k_x' \Delta x)}{\Delta x^2} + \frac{1-\cos(k_y' \Delta y)}{\Delta y^2} + \frac{1-\cos(k_z' \Delta z)}{\Delta z^2}\right) \quad (35)$$

The stability condition for Poisson equation is:

$$\frac{1}{\Delta x^2} + \frac{1}{\Delta y^2} + \frac{1}{\Delta z^2} \leq \frac{1}{4} \quad (36)$$

## 2.1.5 Solving Poisson Equation as An Elliptic PDE

The previous derivation process demonstrates that solving the N-P and Poisson equations using the same numerical method yields two stability conditions, Eqs. (30) and (36). The parameters used in the model must satisfy both criteria simultaneously, significantly restricting the range of suitable values. This poses challenges for modeling realistic electrochemical systems, especially in cases requiring high-voltage batteries or highly concentrated electrolytes. Ignoring these stability conditions can result in anomalous and incorrect concentration changes. To address this issue, the Poisson equation can be treated as an elliptic equation, as stability is generally not a concern in this case due to the absence of time-stepping.

The Eq.(6) can be rewritten as:

$$\frac{\partial^2 \Phi}{\partial x^2} + \frac{\partial^2 \Phi}{\partial y^2} + \frac{\partial^2 \Phi}{\partial z^2} = -\frac{\rho}{\epsilon} \quad (37)$$

where $\rho = F z_\alpha c_\alpha / \epsilon$.

2.1.5.1 Discretizing and Solving

The second derivatives can be approximated:

$$\begin{cases} \frac{\partial^2 \Phi}{\partial x^2} \approx \frac{\Phi_{r+1,j,s} - 2\Phi_{r,j,s} + \Phi_{r-1,j,s}}{\Delta x^2} \\ \frac{\partial^2 \Phi}{\partial y^2} \approx \frac{\Phi_{r,j+1,s} - 2\Phi_{r,j,s} + \Phi_{r,j-1,s}}{\Delta y^2} \\ \frac{\partial^2 \Phi}{\partial z^2} \approx \frac{\Phi_{r,j,s+1} - 2\Phi_{r,j,s} + \Phi_{r,j,s-1}}{\Delta z^2} \end{cases} \quad (38)$$

Then, the discrete form is:

$$\frac{\Phi_{r+1,j,s} - 2\Phi_{r,j,s} + \Phi_{r-1,j,s}}{\Delta x^2} + \frac{\Phi_{r,j+1,s} - 2\Phi_{r,j,s} + \Phi_{r,j-1,s}}{\Delta y^2} + \frac{\Phi_{r,j,s+1} - 2\Phi_{r,j,s} + \Phi_{r,j,s-1}}{\Delta z^2} = -\frac{\rho}{\epsilon}$$

(39)

Rearranging

$$\Phi_{r,j,s} = \frac{1}{2\left(\frac{1}{\Delta x^2} + \frac{1}{\Delta y^2} + \frac{1}{\Delta z^2}\right)} \left( \frac{\Phi_{r+1,j,s} + \Phi_{r-1,j,s}}{\Delta x^2} + \frac{\Phi_{r,j+1,s} + \Phi_{r,j-1,s}}{\Delta y^2} + \frac{\Phi_{r,j,s+1} + \Phi_{r,j,s-1}}{\Delta z^2} + \frac{\rho}{\epsilon} \right)$$

(40)

To simply this equation, we can set $\Delta x = \Delta y = \Delta z = \Delta_{cons}$, which means the grid in domain is uniformly distributed. Then,

$$\Phi_{r,j,s} = \frac{1}{6}\left( \Phi_{r+1,j,s} + \Phi_{r-1,j,s} + \Phi_{r,j+1,s} + \Phi_{r,j-1,s} + \Phi_{r,j,s+1} + \Phi_{r,j,s-1} + \Delta_{cons}^2 \frac{\rho}{\epsilon} \right) \quad (41)$$

The Gauss-Seidel method can be used to solve this equation, and the $\Phi_{r,j,s}$ at each point based on the latest values of its neighboring points.

2.1.5.2 Stability Analysis

The Poisson equation in its standalone form is elliptic, describing the steady-state distribution of electric potential. Since there is no time variable, it doesn't require stability analysis in the usual sense. Instead, we are concerned with convergence and accuracy in solving for the electric potential, such as:

$$\text{diff} = \max | \phi_{r,j,s}^{(new)} - \phi_{r,j,s}^{(old)} | \quad (42)$$

If diff is smaller than a predefined tolerance (i.e. $10^{-6}$), the procedure considers the solution to have converged. If not, the iteration process needs to be repeated.

In another case, if we consider that the N-P equation describes the time-dependent transport of ions, and the Poisson equation determines the electric potential distribution due to the charge density from the ion concentrations. Therefore, in the PNP system, the update for the electric potential (from Poisson's equation) will affect how the concentrations evolve over time, and vice versa. As a result, the entire system, including both the Poisson and N-P equations, needs to be analyzed for stability when using explicit time-stepping schemes.

By combining the amplification factors for both the N-P and Poisson equations, a total amplification factor, $\lambda_{total}$, for the PNP system is obtained:

$$\lambda_{total} = \lambda_{N-P} + \lambda_{\Phi} \quad (43)$$

where $\lambda_{N-P}$ represents the amplification factor of N-P equation. The overall stability condition requires $|\lambda| \leq 1$, leading to constraints on the time step, spatial discretization, and diffusion coefficients, etc.

This stability condition is more restrictive than independently ensuring $|\lambda_{N-P}| \leq 1$ and $|\lambda_{\Phi}| \leq 1$. It is more impossible to simulate successfully a real electrochemical system.

**2.1.6 Boundary Conditions**

As mentioned above, one IC and six BCs are needed for N-P equation, and other six BCs for Poisson equation. In our case, the system has the given boundaries, therefore, the finite layer boundary conditions are considered.

The flux at $x$-, $y$-directions and top of $z$-direction boundaries remains unchanged (Neumann condition for concentration), and the concentrations at the boundaries equal zero:

$$\begin{cases} \left.\dfrac{\partial c_\alpha}{\partial x}\right|_{x=x_{left}} = \left.\dfrac{\partial c_\alpha}{\partial x}\right|_{x=x_{right}} = 0 \\ \left.\dfrac{\partial c_\alpha}{\partial y}\right|_{y=y_{left}} = \left.\dfrac{\partial c_\alpha}{\partial y}\right|_{y=y_{right}} = 0 \quad (44) \\ \left.\dfrac{\partial c_\alpha}{\partial z}\right|_{z=z_{top}} = 0 \end{cases}$$

$$\begin{cases} c_\alpha(x_{left}, y, z, t) = c_\alpha(x_{right}, y, z, t) = 0 \\ c_\alpha(x, y_{left}, z, t) = c_\alpha(x, y_{right}, z, t) = 0 \end{cases} \quad (45)$$

And on the electrode surface, that is the position at $z_{bottom}$, the B-V equation will be obeyed due to the electrochemical reaction:

$$-D_{\alpha_z} \frac{\partial c_\alpha}{\partial z}\bigg|_{z=z_{\text{bottom}}} - \nu_{\alpha_z} c_\alpha \frac{\partial \Phi}{\partial z}\bigg|_{z=z_{\text{bottom}}} = \frac{J_0}{z_\alpha F}\left[\exp\left(\frac{(1-\beta)F\eta}{RT}\right) - \exp\left(\frac{-\beta F\eta}{RT}\right)\right] \quad (46)$$

where $J_0$ is the exchange current density; $\eta = \Phi_{\text{electrode}} - \Phi_{\text{equilibrium}}$ is the overpotential; $\beta$ is the transfer coefficient (typically 0.5 or between 0.3 and 0.7); $\nu_{\alpha_z} = \frac{D_{\alpha_z} z_\alpha F}{RT}$ is the mobility of ion species α ($z_\alpha$ is the charge number of the ion).

The potentiostatic model was used, we have:

$$\Phi(x, y, z_{\text{top}}) = \Phi_{\text{applied}} \quad (47)$$

We can set it as 3 V vs. SHE. At the $z_{\text{bottom}}$, the electric potential also obeys the B-V equation. For x and y- direction, we have:

$$\begin{cases} \frac{\partial \Phi}{\partial x}\bigg|_{x=x_{\text{left}}} = \frac{\partial \Phi}{\partial x}\bigg|_{x=x_{\text{right}}} = 0 \\ \frac{\partial \Phi}{\partial y}\bigg|_{y=y_{\text{left}}} = \frac{\partial \Phi}{\partial y}\bigg|_{y=y_{\text{right}}} = 0 \end{cases} \quad (48)$$

Finally, because we only consider one component α, which means only one ion was studied in this single-ion model. The calculated 3-D concentration profiles are show in Figure 3. All the parameters were set as: $F$=96485 C mol$^{-1}$; $R$=8.314 J mol$^{-1}$ K$^{-1}$; $T$=298 K; $z_\alpha$=1; $D_{\alpha_x} = D_{\alpha_y} = 1\times10^{-5}$ m$^2$ s$^{-1}$; $D_{\alpha_z} = 1\times10^{-2}$ m$^2$ s$^{-1}$; $\epsilon$ =80×8.854e$^{-12}$ F m$^{-1}$; $J_0$=1.0 mA cm$^{-2}$; $\beta$=0.5; $x, y, z \in [0,1]$; $t_F$=5 s.

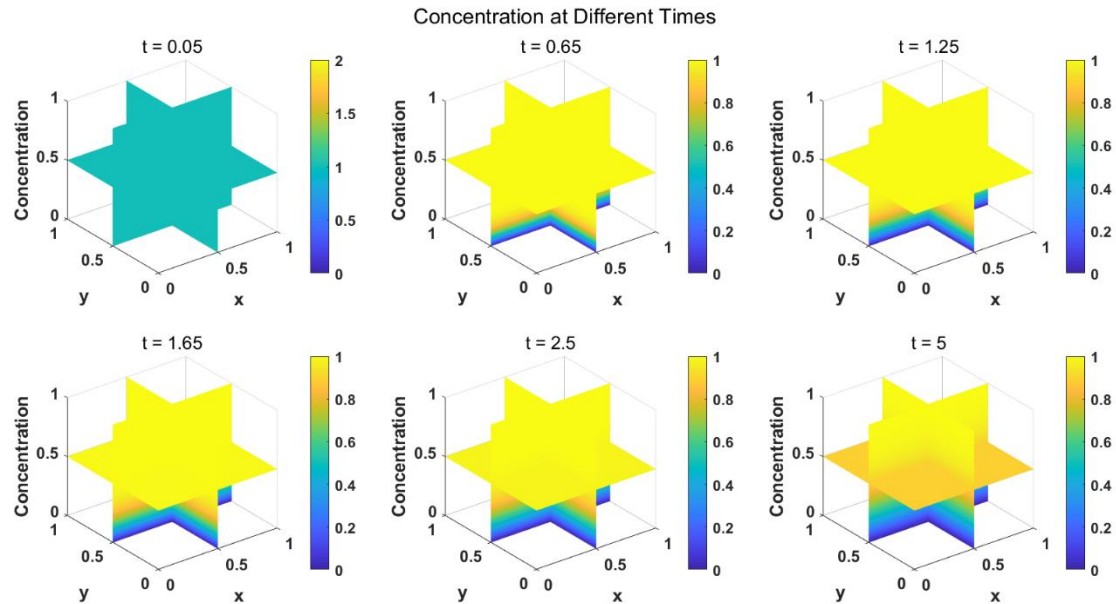

Figure 3. Calculated evolution of 3-D concentration profiles (unit: mol L$^{-1}$) within 5 s under the applied finite layer boundary conditions for a single ion species.

Figure 3 illustrates the evolution of the single-ion system under finite layer boundary conditions, showcasing the interplay of diffusion, drift driven by the electric field, and electrochemical reactions at the electrode surface. Initially, the concentration profile is nearly uniform, representing the even distribution of ions as per the initial conditions. By t=0.65, slight concentration gradients emerge near the boundaries, influenced by flux conditions and the applied potential. Over time, the system stabilizes, characterized by significant ion depletion near the electrode surface due to the electrochemical reactions.

**2.2 Two Ion Species for 1:1 Electrolyte System**

**2.2.1 Basic Model**

To extend the model to account for two ion species in a 1:1 electrolyte system, we need to include the interactions and dynamics of both cation and anion species.
In N-P equation, the flux for each species becomes:

$$\begin{cases} \mathbf{J}_{cat} = -D_{cat}\nabla c_{cat} - \dfrac{D_{cat} z_{cat} F}{RT} c_{cat} \nabla \Phi \\ \mathbf{J}_{an} = -D_{an}\nabla c_{an} - \dfrac{D_{an} z_{an} F}{RT} c_{an} \nabla \Phi \end{cases} \quad (49)$$

where $z_{cat} = 1$ and $z_{an} = -1$. Combined with the continuity equation, we have:

$$\begin{cases} \dfrac{\partial c_{cat}}{\partial t} = -\nabla \cdot \left( D_{cat}\nabla c_{cat} + \dfrac{D_{cat} F}{RT} c_{cat} \nabla \Phi \right) \\ \dfrac{\partial c_{an}}{\partial t} = -\nabla \cdot \left( D_{an}\nabla c_{an} - \dfrac{D_{an} F}{RT} c_{an} \nabla \Phi \right) \end{cases} \quad (50)$$

and

$$\dfrac{\partial c_{total}}{\partial t} = -\nabla \cdot \left[ D_{cat}\nabla c_{cat} + D_{an}\nabla c_{an} + \dfrac{F}{RT}(D_{cat} c_{cat} - D_{an} c_{an})\nabla \Phi \right] \quad (51)$$

The charge density for 1:1 electrolyte is:

$$\rho = F(z_{cat} c_{cat} + z_{an} c_{an}) = F(c_{cat} - c_{an}) \quad (52)$$

Then, the Poisson equation is:

$$\nabla \cdot (\epsilon \nabla \Phi) = -F(c_{cat} - c_{an}) \quad (53)$$

The equations can be similarly extended to 3-D and solved using the FDM. In practice, there is no significant difference between a single-ion system and a two-ion species system when considering the cation, as the boundary condition dictated by the B-V equation applies only to the cation. This highlights that anions are not involved in the electrochemical reactions at the electrode surface. Moreover, the theoretical formulation of the two-ion PNP equations inherently couples the motion of cations and anions through the electric potential, which is governed by the Poisson equation. While these equations do not explicitly include short-range ion-ion interactions, they account for indirect interactions mediated by the shared electric field. Addressing this limitation would require quantum mechanical equations to capture ion-ion interactions, but this would significantly increase the complexity. Furthermore, the current PNP equations in this study are constrained by strict stability criteria, limiting the parameter range for simulations. For instance, Figure 4 illustrates the total cation and anion concentrations under these conditions.

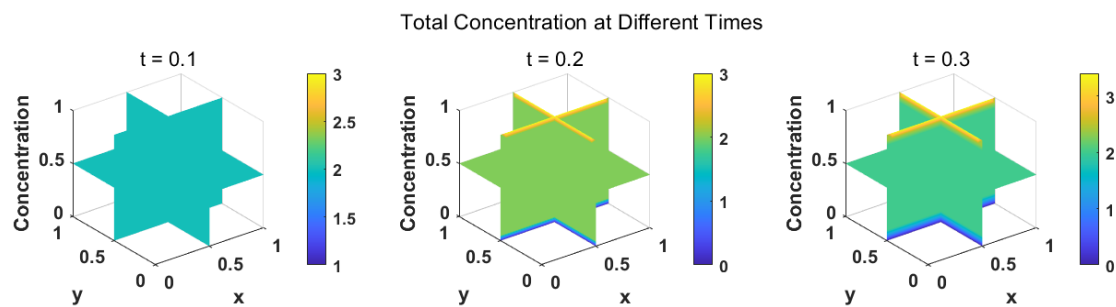

Figure 4. Calculated evolution of 3-D total concentration profiles (unit: mol L$^{-1}$) under the applied finite layer boundary conditions for two ion species.

It can be observed that the electric field and the electrochemical reaction cause a concentration drop near the electrode surface, while ion accumulation occurs at the boundary far from the surface. However, for time intervals exceeding 0.3 seconds, the results become unreliable due to the limitations of the parameter range and the unsatisfied stability criteria. For examples, the available diffusion coefficient should be less than $10^{-2}$ m$^2$ s$^{-1}$. The time step size cannot be less than 50.

**2.2.2 Influence of Spatial and Temporal Parameters**

The influence of number of grid points ($N_x$, $N_y$, and $N_z$) is investigated. The results of different grid points of 10 and 30 were compared, as shown in Figure 5.

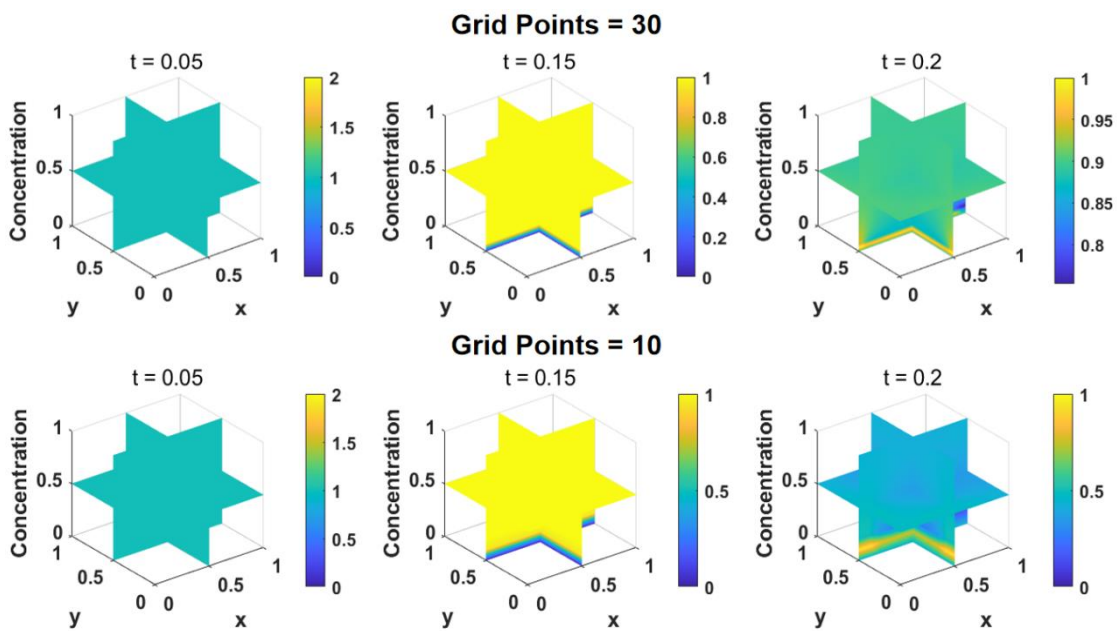

Figure 5. Comparison of calculated concentration profiles (unit: mol L$^{-1}$) of cation derived from the models with different grid points of 10 and 30.

The same parameters are used except for the grid points to investigate its impact. The result highlights the importance of selecting an appropriate grid resolution in numerical simulations of electrochemical systems. Generally, higher grid resolution (N=30) provides more accurate and detailed results, meanwhile, the lower grid resolution (N=10) reduces computational time but sacrifices accuracy, especially in regions with steep gradients near the electrode surface.

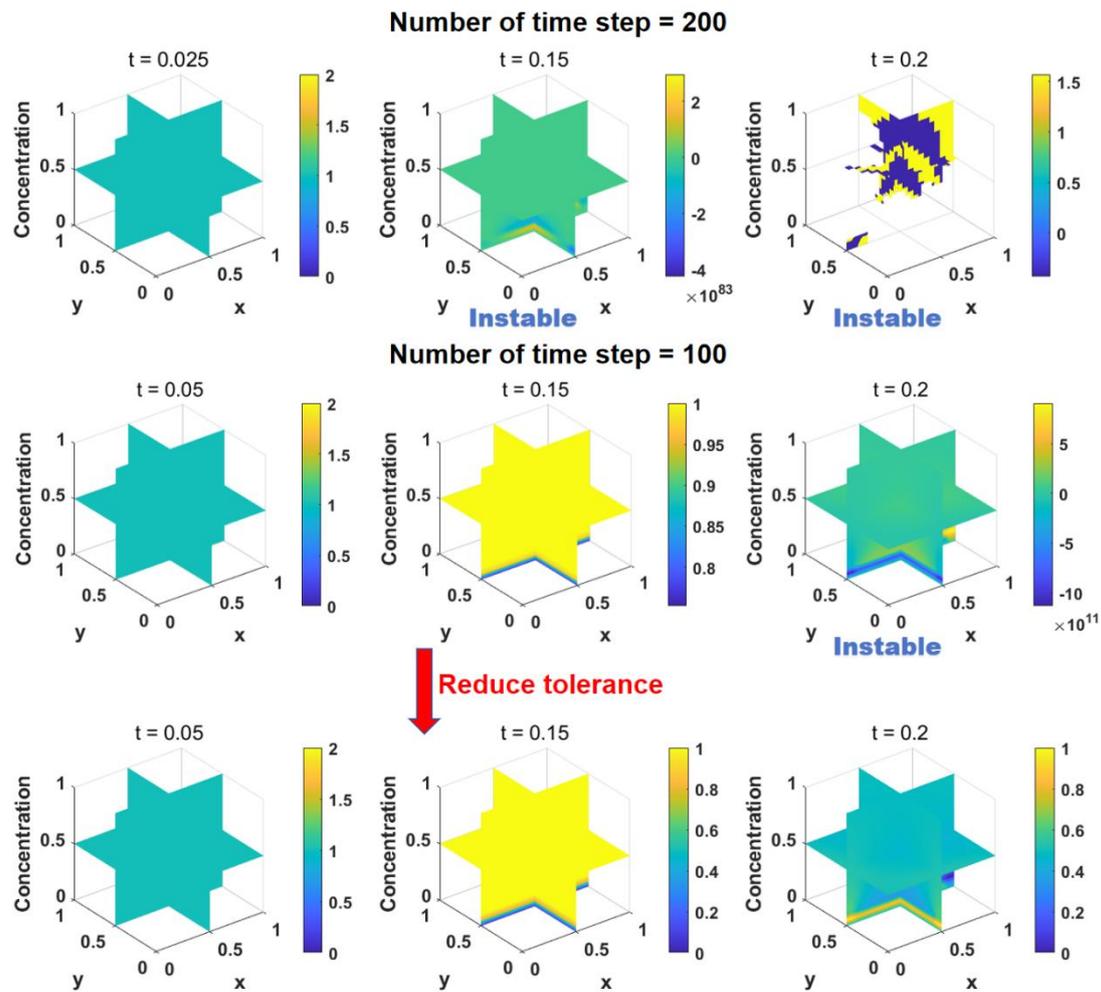

Figure 6. Comparisons of calculated concentration profiles (unit: mol L$^{-1}$) of cation derived from the models with different grid points of 10 and 30.

Figure 6 illustrates the influence of number of time steps (Nt=100 and Nt=200) and numerical tolerance on the stability and accuracy of the simulation results for cation concentration distributions over time. For the top row in the figure, Nt=200, while the smaller time step size initially captures the dynamics with higher resolution, severe numerical instability emerges at the later time, resulting in incomplete data, unphysical spikes and divergence. This instability is likely due to accumulated numerical errors or insufficient control over the coupling between the Nernst-Planck and Poisson equations. At a lower number of time steps (Nt=100), a more stable solution is observed at t=0.05 s and t=0.15 s, but instability still develops at t=0.2 s, suggesting that while reducing the number of time steps mitigates instability, it does not fully resolve it.

In the bottom row in the figure, reducing the numerical tolerance (from $10^{-6}$ to $10^{-4}$) for the convergence criterion improves stability furtherly, as shown by smoother and more realistic concentration profiles. This adjustment ensures that the iterative solver for the Poisson equation achieves better accuracy, thereby reducing the propagation of numerical errors throughout the simulation. The results emphasize the importance of selecting appropriate time steps and tolerances to maintain both stability and accuracy in modeling complex electrochemical systems. We recommend that the optimal range for the number of time steps lies between 50 and 80 to achieve a balance between computational efficiency and numerical stability.

## 3 Conclusions and Perspectives

In this work, we developed a theoretical framework to model concentration distributions in a 3-D electrochemical system by solving the PNP equations using the FDM. Our analysis emphasized the stringent stability conditions required for coupling the Nernst-Planck and Poisson equations across multiple dimensions. These rigorous criteria are particularly necessary when employing explicit numerical schemes to ensure accurate simulation results and to avoid spurious oscillations or numerical artifacts. However, the derived stability conditions impose strict constraints on the choice of time steps and spatial resolution, thereby limiting the range of parameter values that can be effectively simulated. These limitations are most pronounced in the inability of the model to handle high electrolyte concentrations or extreme electric fields, which could lead to unphysical or unstable solutions.

Furthermore, the model applicability is constrained by a short suitable data range, where meaningful results can only be obtained within a limited timeframe before stability issues arise. This becomes particularly problematic for simulating long-term processes or systems requiring high temporal resolution. Additionally, the model does not account for ion-ion interactions, reducing its ability to capture nonlinear effects such as ion crowding or complex electrostatic interactions, which are critical in highly concentrated systems. Despite these challenges, the framework provides valuable insights into the interplay of diffusion, migration, and reaction dynamics, offering a solid foundation for further advancements.

Future work could enhance this model by integrating more complex boundary conditions and additional ion species to better represent real-world electrochemical systems. Furthermore, this work provides an essential resource for developing AI-based models in electrochemical systems. The detailed physical modeling and rigorous stability analysis presented here can serve as training data or benchmarks for machine learning algorithms aiming to predict ion transport and reaction dynamics. The integration of computational techniques with AI could enable the creation of highly efficient predictive models, capable of simulating large-scale or complex electrochemical systems with improved accuracy and reduced computational costs. This symbiotic relationship between traditional modeling and AI has the potential to drive significant advancements in the understanding and design of next-generation electrochemical devices such as batteries and fuel cells.

**Acknowledgement**
The authors gratefully thank the financial support of the National Natural Science Foundation of China (No. 52202200), the Excellent Young Talents Fund Program of Higher Education Institutions of Anhui Province (No. 2022AH030048).